\documentclass[12pt]{iopart}

\begin{document}

\title[Local structure of K$_{0.8}$Fe$_{1.6}$Se$_{2}$ by EXAFS]{Large local disorder in the superconducting
K$_{0.8}$Fe$_{1.6}$Se$_{2}$ studied by extended x-ray absorption fine
structure}

\author{A. Iadecola$^{1}$, B. Joseph$^{1}$, L. Simonelli$^{2}$, A.
Puri$^{3}$, Y. Mizuguchi$^{4,5}$, H. Takeya$^{4,5}$,
Y. Takano$^{4,5}$, N. L. Saini$^{1}$}
\address{$^{1}$Dipartimento di Fisica, Universit\`{a} di Roma ``La
Sapienza", P. le Aldo Moro 2, 00185 Roma, Italy} 
\address{$^{2}$European Synchrotron Radiation Facility, 6 RUE Jules
Horowitz BP 220 38043 Grenoble Cedex 9 France} 
\address{$^{3}$INFN -Laboratori Nazionali di Frascati, via E. Fermi 40, 00044 Frascati,
Roma, Italy}
\address{$^{4}$National Institute for Materials Science, 1-2-1 Sengen, 
Tsukuba 305-0047, Japan}
\address{$^{5}$JST-TRIP, 1-2-1 Sengen,Tsukuba 305-0047, Japan}

\begin{abstract}

We have measured local structure of superconducting
K$_{0.8}$Fe$_{1.6}$Se$_{2}$ chalcogenide (T$_{c}$=31.8 K) by
temperature dependent polarized extended x-ray absorption fine
structure (EXAFS) at the Fe and Se K-edges.  We find that the system
is characterized by a large local disorder.  The Fe-Se and Fe-Fe
distances are found to be shorter than the distances measured by
diffraction, while the corresponding mean square relative displacements 
reveal large Fe-site disorder and relatively large c-axis disorder.  
The local force constant for Fe-Se bondlength (k$\sim$5.8 eV/\AA$^{2}$) is
similar to the one found in the binary FeSe superconductor, however,
the Fe-Fe bondlength appears to get flexible (k$\sim$2.1 eV/\AA$^{2}$)
in comparison to the binary FeSe (k$\sim$3.5 eV/\AA$^{2}$), an indication
of partly relaxed Fe-Fe networks in K$_{0.8}$Fe$_{1.6}$Se$_{2}$.
The results suggest glassy nature of the title system, with the
superconductivity being similar to the one in the granular materials.\\

\noindent Journal reference: Journal of Physics: Condensed Matter 24 (2012) 115701

\end{abstract}


\maketitle

\section{Introduction}

Discovery of superconductivity in the iron-based `1111' pnictides
\cite{Kamihara} boosted renewal of research activities in the field,
resulting a number of iron containing superconductors with different
transition temperatures, the maximum being about 55 K for the SmFeAsO
system \cite{Johnston,TakanoRev}.  Among these, the FeSe (the 11-type
chalcogenide) shows lowest superconducting transition temperature
(T$_{c}\sim$8 K), however, could be considered as a model system to
address basic characteristics of these materials \cite{TakanoRev}.
The FeSe structure contains simple stacking of tetrahedrally
coordinated FeSe$_{4}$ layers without spacer layers that are known to
have substantial effect on the electronic properties
\cite{IadecolaEPL09,JosephJPCM}.  Susbstitution by Te in the FeSe
leads to a marginal increase in the T$_{c}$ ($\sim$ 15 K), however,
the system gets phase separated and the nanoscale structure is
characterized by different iron-chalcogen bondlengths
\cite{JosephPRB}.  On the other hand, the superconducting transition
temperature of the FeSe shows large enhancement up to $\sim$37 K under
the hydrostatic pressure \cite{Mizuguchi,Medvedev}.  A large pressure
sensitivity of the FeSe indicates chemical pressure being a potential
alternative parameter for raising its T$_{c}$.  Indeed,
superconductivity at a T$_{c}$ as high as 32 K has been observed
recently in a K-intercalated FeSe \cite{Guo}, triggering new studies on
the iron-based chalcogenides.  Later, similar T$_{c}$ has been
observed in the Rb$_{x}$Fe$_{2-y}$Se$_{2}$ \cite{WangPRB,LiPRB11},
(Tl,K)Fe$_{2-y}$Se$_{2}$ \cite{FangEPL} and Cs$_{x}$Fe$_{2-y}$Se$_{2}$ \cite{YingPRB} compounds.

There appears a substantial charge transfer between the active FeSe
blocks and the intercalated spacer layers in the
K$_{x}$Fe$_{2-y}$Se$_{2}$, with electrons from the K$^{+}$ ions
filling the hole bands at the zone center.  Thus, the Fermi surface of
K$_{x}$Fe$_{2-y}$Se$_{2}$ is made by only the electron pockets around
the M points \cite{QianPRL,ZhaoPRB}, and hence the scenario based on
the nesting between electron and hole pockets, usually argued to
understand the iron pnictides, looses its ground \cite{Mazin11}.  On
the other hand, although isostructural to the 122-type
BaFe$_{2}$As$_{2}$ pnictides, the K$_{x}$Fe$_{2-y}$Se$_{2}$ has
distinct microstructure, characterized by an iron-vacancy order and a
phase separation \cite{Ricci11,YePRL,ZavalijPRB,WangPRBR}.  The
magnetic order with a remarkably large iron magnetic moment co-exists with bulk superconductivity \cite{BaoCPL,Pomjakushin}.  In
addition, there are experimental indications on the relation between
the superconductivity and the iron-vacancy disorder, with a completely
ordered state being an insulator.  These facts point out that a 
complete knowledge of the nanoscale structure is needed for describing the
fundamental electronic structure of these materials.

Extended x-ray absorption fine structure (EXAFS), an atomic
site-specific experimental probe \cite{Konings}, has been widely used
to study local structure of different materials, including the copper
oxide superconductors and related systems \cite{StrBonding}.  EXAFS has
also been exploited to study the new iron-based superconductors,
revealing important information on the atomic correlations between
different layers \cite{IadecolaEPL09,JosephJPCM} and temperature
dependent local structural anomalies across the superconducting
transition \cite{OyanagiPRB,JosephNdJPCM}.  More recently, we have
applied EXAFS to study the local structure of FeSe$_{1-x}$Te$_{x}$
chalcogenides, providing direct evidence of phase separation in the
ternary systems, characterized by different Fe-Se and Fe-Te distances
\cite{JosephPRB}.  In the present study, we have used polarized EXAFS
to explore the local structure of K$_{0.8}$Fe$_{1.6}$Se$_{2}$
superconductor.  The results of polarized EXAFS at the Fe and Se
K-edges reveal large local disorder similar to the amorphous/glass
materials.  The local structure of K$_{0.8}$Fe$_{1.6}$Se$_{2}$ is
significantly different from the local structure of the binary FeSe
\cite{JosephPRB} superconductor.  While the local force constant of
Fe-Se bondlength (k$\sim$5.8 eV/\AA$^{2}$) is similar in the two
systems, the one for the Fe-Fe bondlength appears to sustain
substantial changes due to K intercalation (k$\sim$3.5 eV/\AA$^{2}$
and k$\sim$2.1 eV/\AA$^{2}$ respectively for FeSe and
K$_{0.8}$Fe$_{1.6}$Se$_{2}$ systems), consistent with a partial
relaxation of the Fe-Fe networks due to the K-intercalation.  The
experimental results suggest that glass physics should have distinct
role in the understanding of these superconductors.

\section{Experimental Details}

X-ray absorption measurements were performed at the beamline BM23 in
the European Synchrotron Radiation Facility (ESRF), Grenoble on well
characterized single crystal of K$_{0.8}$Fe$_{1.6}$Se$_{2}$ system.
The single crystal was grown by melting a precursor of FeSe and K
placed in an alumina crucible, that itself was sealed into an
arc-welded stainless steel tube.  The details on the crystal growth
and characterization for superconducting, structural and transport
properties are described elsewhere \cite{TakanoAPL}.  A platelike and
dark-shining as grown single crystal with T$_{c}$=31.8 K was selected
and detailed synchrotron x-ray diffraction study \cite{Ricci11}
was performed prior to the x-ray absorption measurements.  The
synchrotron radiation emitted by a bending magnet source at 6 GeV ESRF
storage ring was monochromatized by a double crystal Si(311)
monochromator.  The polarized EXAFS measurements at the Fe and Se
K-edges were acquired in the normal incidence (with polarization
parallel to the ab-plane and hereafter called E$\parallel$ab
polarization) and in the grazing incidence geometry (10 degree off the
E$\parallel$c and hereafter called E$\parallel$c polarization).
Fluorescence mode was used (exploiting a multi-element Ge-detector
system) for the measurements.  The samples were mounted in a
continuous flow He cryostat and the temperature was controlled and
monitored within an accurancy of $\pm$1 K. Several spectral scans were
acquired at each temperature to ensure the reproducibility of the
measurements.  Standard procedure was used to extract the EXAFS signal
from the absorption spectrum \cite{Konings} and corrected for the
fluorescence self-absorption before the analysis \cite{Troger}.

\section{Results and Discussion}

%Figure 1
\begin{figure}
\input{epsf}
\epsfxsize 12.5cm
\centerline{\epsfbox{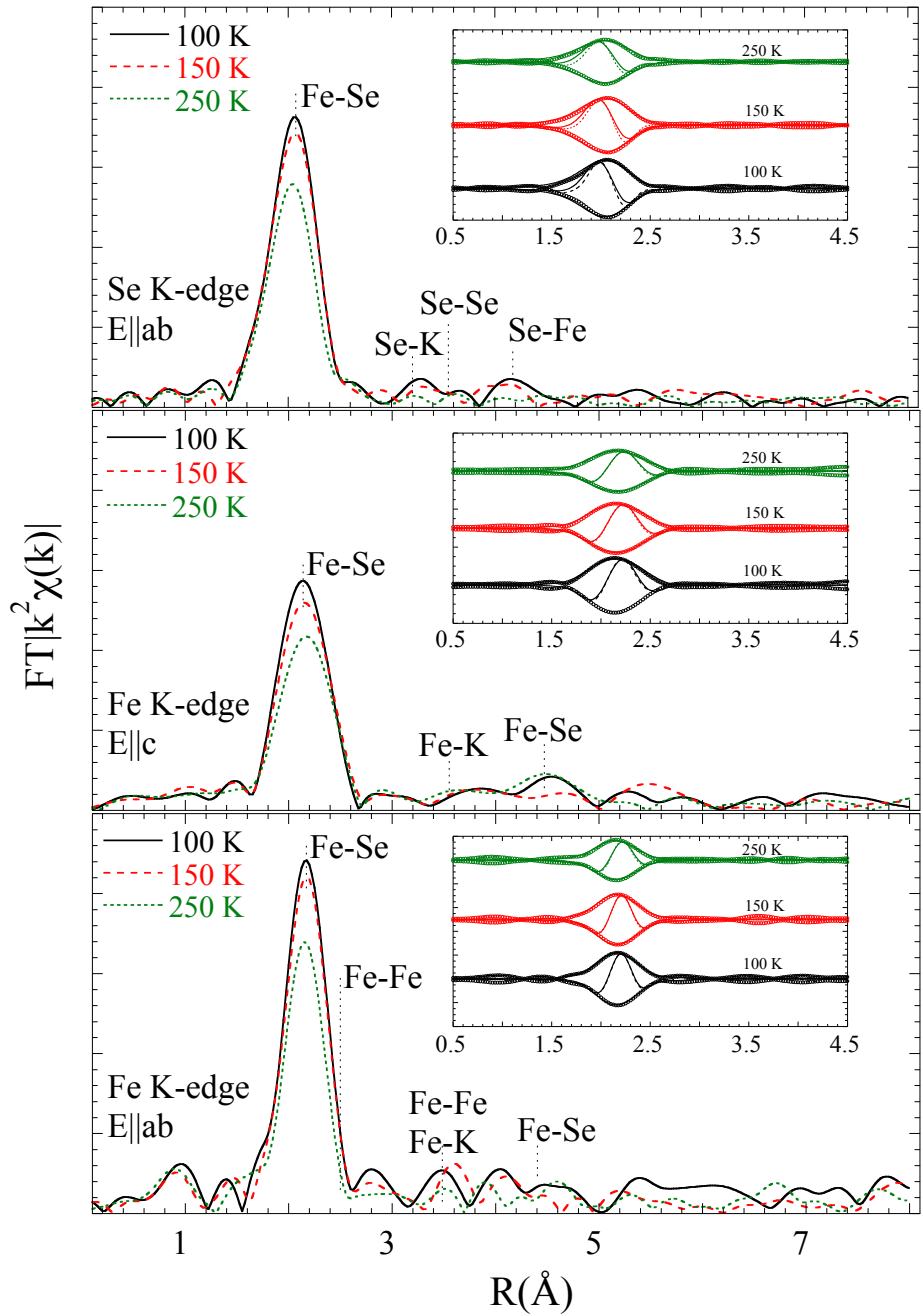}}
\caption{\label{fig:epsart}
Fourier transform (FT) magnitudes of the polarized EXAFS (weighted by
k$^{2}$) measured on K$_{0.8}$Fe$_{1.6}$Se$_{2}$ single crystal at
several temperatures.  The measurements were made at Se K-edge in the
E$\parallel$ab (upper) polarization and, at Fe K-edge in the
E$\parallel$ab (lower) and E$\parallel$c (middle) polarizations.  The
FT are performed using a Gaussian window and not corrected for the
phase shifts.  The k-range is 3.5-15.5 \AA$^{-1}$ for the
E$\parallel$ab EXAFS, and 3.5-14 \AA$^{-1}$ for the E$\parallel$c data.
The insets show experimental FT magnitudes and real parts with the
model fits (solid lines).}
\end{figure}

Figure 1 shows Fourier transform (FT) magnitudes of the polarized
EXAFS oscillations, measured at the Se and Fe K-edges on
K$_{0.8}$Fe$_{1.6}$Se$_{2}$ single crystal.  The FT magnitudes provide
partial atomic distribution around the Se and Fe in the direction of
x-ray beam polarizations.  Se K-edge EXAFS, measured in the
E$\parallel$ab polarization shows only the contribution from the
nearest neighbour Fe atoms at a distance $\sim$2.4 \AA$ $ (main peak in 
the FT at $\sim$2 \AA$ $).  The next nearest neighbours of Se in the
E$\parallel$ab geometry are K (R=3.45 \AA), Se (R=3.91 and 4.02 \AA) and
Fe (R=4.62 \AA) atoms, and their contributions are expected to appear
at $\sim$3.0-4.5 \AA$ $ (upper panel).  On the other hand, the Fe site
nearest neighbours are Se (at a distance $\sim$2.4 \AA) and Fe (at a
distance $\sim$ 2.7 \AA) in the E$\parallel$ab polarization (lower
panel).  The contribution of the Fe-Fe is not expected in the
E$\parallel$c polarization and the main peak in the FT in this geometry
should be merely due to Fe-Se bonds (middle panel).  Contributions of
Fe next near neighbours are expected to appear at longer distances
(R$_{Fe-Fe}$=3.91 \AA$ $, R$_{Fe-K}$=4.02 \AA$ $ and R$_{Fe-Se}$=4.62
\AA$ $).  Therefore, the main peak in the FT of the Fe K-edge EXAFS in the
E$\parallel$ab geometry contains information on the Fe-Se and Fe-Fe
bonds (lower panel) while the one in the E$\parallel$c have
contribution only of Fe-Se bonds (middle panel), similar to the main
FT peak of the Se K-edge EXAFS (upper panel).  As a matter of fact,
apart from the nearest neighbours, contribution expected from the
farther atoms is apparently absent (or within the noise level).  This
is a characteristic feature of highly disordered systems and amorphous
materials \cite{Greaves}.  We will come back to discuss this, however,
before that let us make a comparison between the local structure of
FeSe and the K$_{0.8}$Fe$_{1.6}$Se$_{2}$.

%Figure 2
\begin{figure}
\input{epsf}
\epsfxsize 10.5cm
\centerline{\epsfbox{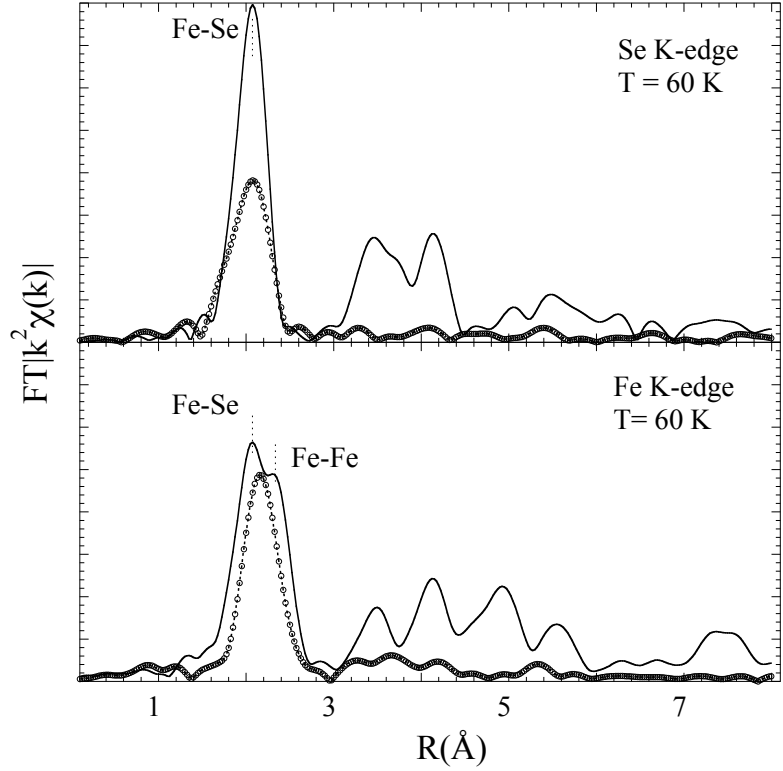}}
\caption{\label{fig:epsart}
Fourier transform magnitudes of the Fe K and Se K-edge EXAFS (weighted
by k$^{2}$) measured on polycrystalline FeSe (solid line) and the
K$_{0.8}$Fe$_{1.6}$Se$_{2}$ crystal (open symbols) in the
E$\parallel$ab geometry at T=60 K. Similar k-range (3.5-15.5
\AA$^{-1}$) is used to perform the FT for the comparison.}
\end{figure}

Figure 2 compares Fourier transforms (FT) of the Fe and Se K-edge EXAFS
signals (T=60 K) measured on the polycrystalline FeSe
sample \cite{JosephPRB} and the K$_{0.8}$Fe$_{1.6}$Se$_{2}$ crystal in
the E$\parallel$ab geometry.  The Fe nearest neighbour in the FeSe are
Se at a distance of $\sim$2.4 \AA$ $ and Fe atoms at a distance $\sim$ 2.7
\AA$ $ (two peak structure at $\sim$1.5-3.0\AA$ $), similar to the case
of K$_{0.8}$Fe$_{1.6}$Se$_{2}$.  Significantly intense features
between 3-6 $\AA$ are due to longer Fe-Fe ($\sim$3.8 $\AA$) and Fe-Se
distances, and the multiple scatterings involving different near
neighbours of the Fe atoms.  Similarly, in the Se K-edge EXAFS of the
FeSe we expect contribution from four Fe nearest neighbours at a
distance $\sim$2.4 \AA$ $ (as in the K$_{0.8}$Fe$_{1.6}$Se$_{2}$).  The
next nearest neighbours are Se and Fe atoms and contributions of these
distant shells clearly appear as a multiple structured peak at
$\sim$3.0-4.5 \AA$ $.

The comparison underlines significant differences in the local
structure of the FeSe and the K$_{0.8}$Fe$_{1.6}$Se$_{2}$ systems.  In
particular, the FT of the K$_{0.8}$Fe$_{1.6}$Se$_{2}$ show only a
single peak at $\sim$2 \AA$ $ that contains contribution from the
Fe-Se distances ($\sim$2.4 \AA) in the Se K-edge EXAFS and, Fe-Se
and Fe-Fe contributions in the Fe K-edge EXAFS. The longer distances
contributions are apparently absent in the K$_{0.8}$Fe$_{1.6}$Se$_{2}$
system.  Also, the FT magnitude due to similar nearest neighbours is
much weaker in the K$_{0.8}$Fe$_{1.6}$Se$_{2}$.  In addition, the
contribution from the Fe-Fe distances in the Fe K-edge EXAFS is
strongly damped, indicating a large Fe site disorder.  Apparently the
EXAFS data reveal large overall local disorder in the
K$_{0.8}$Fe$_{1.6}$Se$_{2}$ system, commonly seen in the local
structure of amorphous systems \cite{Greaves}.  On the other hand, the
measured K$_{0.8}$Fe$_{1.6}$Se$_{2}$ sample is a very good single
crystal, evident from sharp diffraction peaks \cite{Ricci11}, and
such a large disorder in the local structure indicates that the system
should be a some kind of glass.

Coming back, it is clear that only the nearest neighbours (i.e., Fe-Se
and Fe-Fe bondlengths) contributions are visible in both Fe and Se
K-edge EXAFS. Therefore our focus should be limited to these
bondlengths.  In the single-scattering approximation, the EXAFS
amplitude is described by the following general
equation \cite{Konings}:
\begin{equation}
\chi(k)= \sum_{i}\frac{N_{i}S_{0}^{2}}{kR_{i}^{2}}f_{i}(k,R_{i})
e^{-\frac{2R_{i}}{\lambda}} e^{-2k^{2}\sigma_{i}^{2}}
sin[2kR_{i}+\delta_{i}(k)]\nonumber
\end{equation}
where N$_{i}$ = 3 cos$^{2}$ $\theta_{i}$ is the number of neighbouring
atoms at a distance R$_{i}$ at an angle $\theta_{i}$ with respect to
the x-ray beam polarization.  S$_{0}^{2}$ is the passive electrons
reduction factor, f$_{i}$(k,R$_{i}$) is the backscattering amplitude,
$\lambda$ is the photoelectron mean free path, $\delta_{i}$ is the
phase shift, and $\sigma_{i}^{2}$ is the correlated Debye-Waller
factor (DWF) measuring the mean square relative displacements (MSRDs)
of the photoabsorber-backscatterer pairs.  EXCURVE 9.275 code (with
calculated backscattering amplitudes and phase shift functions) is
used for the EXAFS model fits \cite{excurve}.  Similar results were
obtained with WINXAS package \cite{winxas} in which the backscattering
amplitudes and phase shifts were calculated using FEFF \cite{Feff}.
The E$_{0}$, N$_{i}$ and S$_{0}^{2}$ were fixed after a number of
trials on different EXAFS scans with structural input from diffraction
studies \cite{Ricci11,YePRL,ZavalijPRB,WangPRBR,KFeSeStr}.  Only the
radial distances R$_{i}$ and the corresponding $\sigma_{i}^{2}$, were
allowed to vary in the least squares model fits for the temperature
dependent study.  A single shell (i.e., Fe-Se) was used to model the
E$\parallel$ab polarized Se K-edge EXAFS and E$\parallel$c polarized
Fe K-edge EXAFS. On the other hand, two shells (i.e., Fe-Se and Fe-Fe)
were necessary to model the E$\parallel$ab polarized Fe K-edge EXAFS.
The number of independent data points,
N$_{ind}\sim$(2$\Delta$k$\Delta$R)/$\pi$ \cite{Konings} were about
17-18 ($\Delta$k=11-12 \AA$^{-1}$ and $\Delta$R=2.4 \AA), in the two
parameters (four parameters) fits for the E$\parallel$ab Se K-edge and
E$\parallel$c Fe K-edge EXAFS (E$\parallel$ab Fe K-edge EXAFS).  The
model fits are shown as insets to the Fig.  1.

%Figure 3
\begin{figure}
\input{epsf}
\epsfxsize 8.0cm
\centerline{\epsfbox{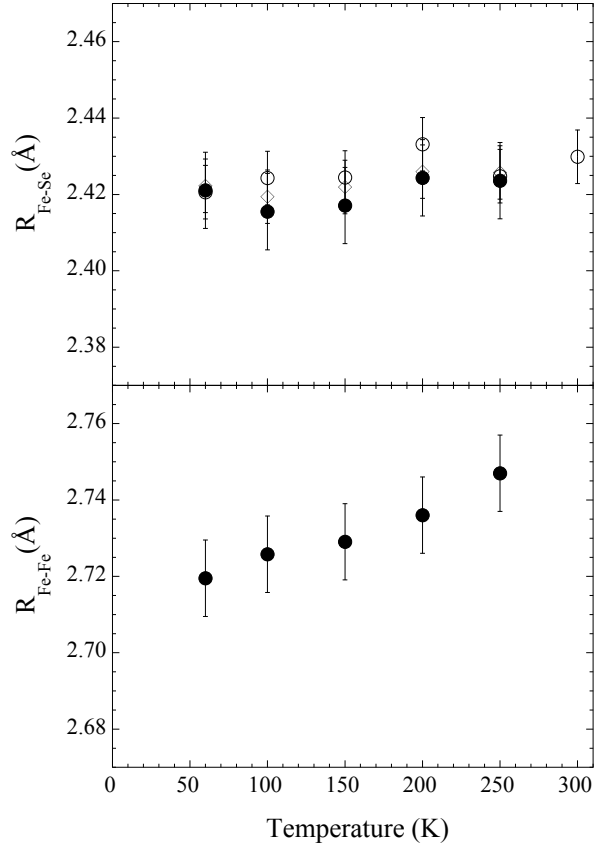}}
\caption{\label{fig:epsart} 
Fe-Se and Fe-Fe distances in the K$_{0.8}$Fe$_{1.6}$Se$_{2}$ as a
function of temperature extracted from the polarized EXAFS spectra
measured at Se and Fe K-edges.  The Fe-Se bondlengths are similar
irrespective of the selected absorbing atom and polarization, and
slightly smaller than the average Fe-Se distance measured by
diffraction.  The open circles are from Se K-edge while the open
diamonds represent the data from the E$\parallel$c polarized Fe K-edge
EXAFS. The closed circles are from the E$\parallel$ab polarized Fe
K-edge EXAFS. The error bars represent maximum uncertainty, determined
using correlation maps between different parameters.}
\end{figure}

Figure 3 shows temperature dependence of the local bondlengths
determined by Fe K and Se K-edge EXAFS analysis.  The Fe-Se distances
determined by EXAFS data in different polarizations on two edges are
similar within experimental uncertainties (upper panel).  However,
both Fe-Se and Fe-Fe distances (lower panel) appear slightly shorter
than the average bondlengths determined by diffraction
studies \cite{Ricci11,YePRL,ZavalijPRB,WangPRBR,KFeSeStr}.  This may be
due to largely disordered structure, as seen in the
glasses \cite{Kolobov,Majid}.  The Fe-Se bondlength hardly shows any
change with temperature unlike the Fe-Fe distance (describing the
in-plane lattice parameter) that varies with temperature showing
relative flexibility of the later.

%Figure 4
\begin{figure}
\input{epsf}
\epsfxsize 9.0cm
\centerline{\epsfbox{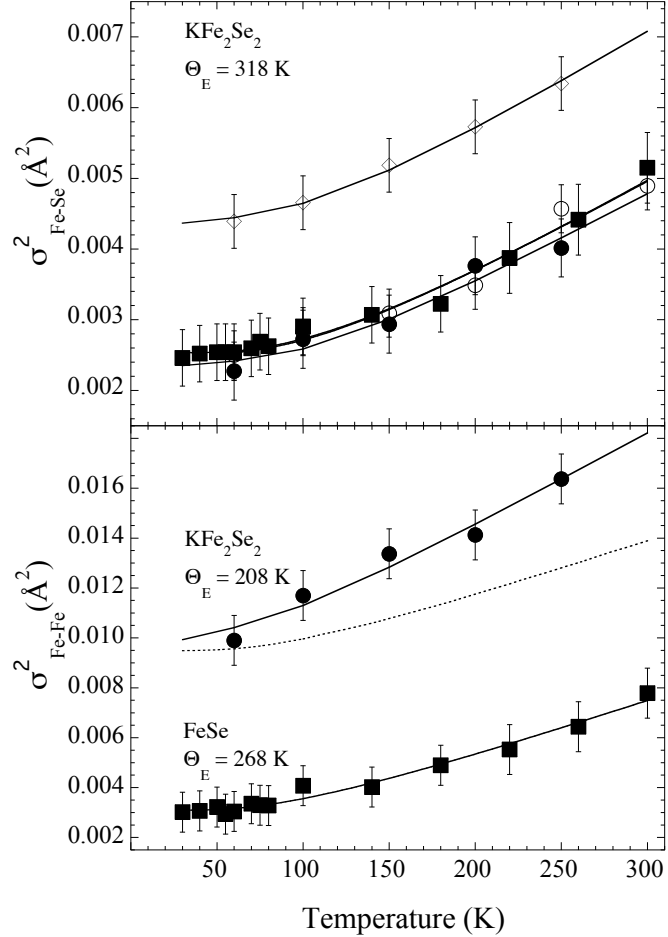}}
\caption{\label{fig:epsart} 
Temperature dependence of the Fe-Se and Fe-Fe MSRDs (symbols) for the
K$_{0.8}$Fe$_{1.6}$Se$_{2}$ system determined by polarized EXAFS
measured at Se and Fe K-edges.  The empty (filled) circles are from
E$\parallel$ab polarized Se K-edge (Fe K-edge) EXAFS while the open
diamonds represent the data from the E$\parallel$c polarized Fe
K-edge.  The filled squares are the data obtained on the binary FeSe
system \cite{JosephPRB}.  The solid lines represent the correlated
Einstein model fit to the data.  The Einstein temperatures,
$\theta_{E}$, are $\sim$318 K for the K$_{0.8}$Fe$_{1.6}$Se$_{2}$,
similar to the Fe-Se bonds in the binary FeSe system (the maximum
uncertainty is about $\pm$20 K).  The $\sigma_{0}^{2}$ in the
E$\parallel$ab polarization is smaller ($\sim$0.0002) in compared to
the one in the E$\parallel$c ($\sim$0.002), consistent with large
static disorder in the c-direction.  While the $\theta_{E}$ for the
Fe-Se remains similar to the binary FeSe system, the $\theta_{E}$ for
the Fe-Fe is significantly different, suggesting large change in these
bonds due to K-intercalation.}
\end{figure}

Figure 4 shows the correlated Debye Waller factors ($\sigma^{2}$),
i.e., the MSRDs of the Fe-Se and the Fe-Fe pairs as a function of
temperature for the K$_{0.8}$Fe$_{1.6}$Se$_{2}$, compared with the
$\sigma^{2}$ of the two bondlengths for the binary FeSe
system \cite{JosephPRB}.  The $\sigma^{2}$ measured by EXAFS
(represents distance broadening) is sum of temperature independent
($\sigma_{0}^{2}$) and temperature dependent terms, i.e.
$\sigma^{2}$=$\sigma_{0}^{2}$+$\sigma^{2}$(T).  While the
$\sigma_{0}^{2}$ ($\sim$0.0002) for the Fe-Se pairs in the
E$\parallel$ab polarization remains small (approximately similar to
the case of the one in the binary FeSe), the $\sigma_{0}^{2}$
($\sim$0.002) for the E$\parallel$c is quite large.  This merely
indicates large disorder along the c-direction, likely to be due to
the K-intercalation.  Also, the $\sigma_{0}^{2}$ for the Fe-Fe is
substantially large, a mere indicator of Fe site disorder.

On the other hand, temperature dependent $\sigma^{2}$(T) is well
described by the correlated Einstein model \cite{Konings,StrBonding},
an appropriate approximation for local mode vibrations.  The Einstein
equation is given as:
\begin{equation}
\sigma^{2}(T)= \frac{\hbar^{2}}{2\mu k_{B}\theta_{E}}coth
\frac{\theta_{E}}{2T}\nonumber
\end{equation}
The Einstein temperature ($\theta_{E}$, i.e., the Einstein frequency
$\omega_{E}$ =$k_{B}\theta_{E}$/$\hbar$) for the Fe-Se pairs remains
similar to the one in the binary FeSe system ($\theta_{E}\sim$318 K),
indicating that local force constant (k=$\mu\omega_{E}^2$, where k is
the effective force constant and $\mu$ is reduced mass of the Fe-Se
pair) for the Fe-Se bonds is not very sensitive to the K-
intercalation between the FeSe layers.  The local force constant for
Fe-Se bondlength is found to be k$\sim$5.8 eV/$\AA^{2}$, similar to
the one in the binary FeSe superconductor.  On the other hand, there
is a substantial effect of intercalation on the Einstein temperature
(local force constant) of the Fe-Fe pairs, estimated to be $\sim$208 K
(k$\sim$2.1 eV/\AA$^{2}$) and $\sim$268 K (k$\sim$3.5 eV/\AA$^{2}$)
respectively for the K$_{0.8}$Fe$_{1.6}$Se$_{2}$ and the binary FeSe.  Therefore, it appears that the Fe-Fe bondlengths are
relatively relaxed in the K$_{0.8}$Fe$_{1.6}$Se$_{2}$,
consistent with a glassy nature of this system.

Considering the fact that the system is having a well defined crystal
ordering, the glassy local structure could be due to freezing of iron
vacancy order coupled with some other degrees of freedom.  Several
measurements have undelined that the magnetic ordering in this system
co-exists with the superconductivity with a large iron moments (3.3
$\mu_{B}$/Fe) ordered antiferromagnetically along the
c-axis \cite{BaoCPL,Pomjakushin}.  Present measurements reveal that
there is; i) a large static disorder along the c-axis and ii) a 
large Fe-site disorder.  In addition, the Fe-Fe network (in-plane
lattice parameter) is relatively relaxed.  Therefore it appears
plausible to think that a frozen state of c-axis disorder (associated
with local strain fields due to K-intercalation and iron vacancy
order) coupled with magnetic order is realized.  This results into a
magnetic texture with large frozen magnetic moment and a relaxed Fe-Fe
network.  Thus, it is likely that the strain fields locally compress
the FeSe-block and hence the T$_{c}$ reaching to a value as high as 32
K, as it happens for FeSe under external pressure.  Indeed,
single-crystal x-ray diffraction has shown a clear evidence of
nanoscale phase separation between a majority magnetic phase with iron
vacancy ordering, co-existing with a minority compressed
phase \cite{Ricci11}.  This situation is similar to granular
superconductors in which a nanoscale superconducting phase coexisting
in an insulating texture.  Therefore, the physics of the
K$_{0.8}$Fe$_{1.6}$Se$_{2}$ system should be quite similar to the
physics of glasses and granular superconductors.  It should be
recalled that the thermodynamic and kinetic fragility of a glass is
related to the number of potential energy minima in the phase space
and the heights of the activation energy barriers separating these
minima.  The fact that the superconductivity in
K$_{0.8}$Fe$_{1.6}$Se$_{2}$ system is strongly dependent on its
thermal history \cite{FeiHan}, a characteristic feature of glasses,
further underlines glassy nature of this system.  The observation of
glassy nature puts the system in the category of granular
superconductors, consistent with the superconductivity in A15
systems \cite{Testardi} and recently observed fractal distribution
enhanced superconductivity in superoxygenated
La$_{2}$CuO$_{4}$ \cite{FratiniNature}.

\section{Summary}

In summary, we have studied local structure of the superconducting
K$_{0.8}$Fe$_{1.6}$Se$_{2}$ chalcogenide by polarized Fe and Se
K-edges EXAFS. The EXAFS data provide a clear evidence of large local
disorder in this system.  Indeed the local structure of K$_(0.8)$Fe$_(1.6)$Se$_2$ is similar to that of the amorphous materials indicating glassy nature of
the system.  The local bondlengths are found to be slightly shorter
than the average diffraction distances, a characteristic feature of
glasses.  The mean square relative displacements (MSRD) of Fe-Se and
Fe-Fe bondlengths are well described by the correlated-Einstein model
with similar Einstein-temperatures in all the polarizations, albeit
the stastic component along the c-axis is much larger, likely to be due
to substitutional disorder in the K layer.  While the Fe-Se
bondlengths remains highly covalent in the
K$_{0.8}$Fe$_{1.6}$Se$_{2}$, similar to the one found in the binary
FeSe superconductor, the Fe-Fe bondlength is charcaterized by much
smaller force constant compared to the binary FeSe.  Such a local
relaxation of the Fe-Fe bondlength results in a compression of the
FeSe unit, as happens under external pressure, and hence the
superconductivity at high T$_{c}$ in the title system should be due to
locally compressed nanoscale minority phase, coexisting with the normal
magnetic phase, similar to the case of granular superconductors.\\
\noindent
{\small {\bf Note added}: After finishing this manuscript we came across a very recent publication by Tyson {\it et al} \cite{Tyson} reporting similar kind
of measurements on the K$_{0.8}$Fe$_{1.6+x}$Se$_{2}$ system finding a
large disorder, similar to the one reported in this manuscript.
However, Tyson {\it et al} have compared the data with the one obtained on
1111-iron pnictide unlike the present study in which a direct
comparison with the binary FeSe has allowed us to clearly identify the
effect on the local structure due to K-intercalation.  Therefore, the
glassy nature revealed in these measurements should provide further
insight to our understanding of these superconductors.}

\section*{Acknowledgments}

The authors wishes to thank ESRF staff for the help and support 
during the experimental runs.

\end{document}